\documentclass[showpacs,aps,prb,twocolumn]{revtex4}
\usepackage{amsmath}
\usepackage{dcolumn}
\usepackage{epsfig}

\def\be{{\bar\eta}}
\def\sg{{\sigma}}

\begin{document}


\title{
First-principles study of epitaxial strain in perovskites
}

\author{Oswaldo Di\'eguez}
\author{Karin M. Rabe}
\author{David Vanderbilt}

\affiliation{Department of Physics and Astronomy, Rutgers University,
             Piscataway, New Jersey 08854-8019, USA}


\begin{abstract}
Using an extension of
a first-principles method developed by King-Smith and
Vanderbilt [Phys.~Rev.~B {\bf 49}, 5828 (1994)], we investigate the
effects of in-plane epitaxial strain on the ground-state structure
and polarization of eight perovskite oxides: BaTiO$_3$, SrTiO$_3$,
CaTiO$_3$, KNbO$_3$, NaNbO$_3$, PbTiO$_3$, PbZrO$_3$, and
BaZrO$_3$.  In addition, we investigate the effects of a nonzero
normal stress. The results are shown to be useful in predicting
the structure and polarization of perovskite oxide thin films and
superlattices.
\end{abstract}

\date{\today}


\maketitle


\section{Introduction}

\begin{figure}
\centerline{\epsfig{file=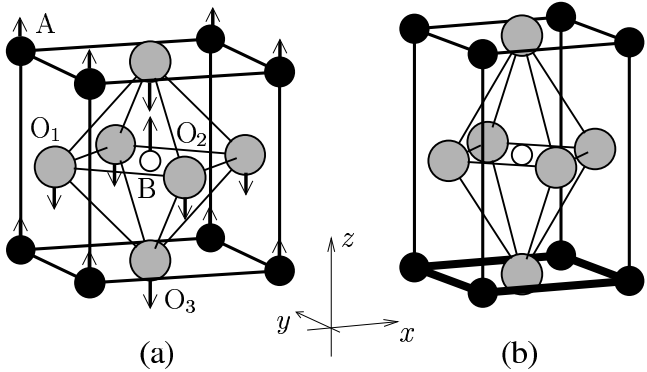,width=3.35in}}
\caption{(a) Ideal cubic perovskite structure for an ABO$_3$ compound;
             the $z$-polarized soft-mode atomic displacements are indicated
             by arrows.
         (b) Epitaxial paraelectric (or $p$) phase, in which the atoms are
             constrained in the $xy$ plane due to the presence of the
             substrate.}
\label{fig_struct}
\end{figure}

Ferroelectrics are insulating solids of technological importance
because of their ability to maintain an electric polarization that
can be reoriented by the application of an electric
field.\cite{Lines1977book} This property lends itself to technological
applications including microelectronic devices and computer
memories.  Among ferroelectrics, perovskites constitute a subclass
that has been of theoretical and experimental interest since the
discovery in 1945 of its first member, barium titanate
(BaTiO$_3$).  This interest is motivated in part by the relative
simplicity of their cubic crystalline phase.  For a perovskite of
general formula ABO$_3$, this structure contains cations A at the
cube corners, a cation B at the center of the cube, and oxygen
atoms at the center of the cube faces forming a regular octahedron,
as depicted in Fig.~\ref{fig_struct}(a).  Typically, perovskites
are found in this cubic paraelectric phase at high temperature; as
the temperature is reduced, symmetry-lowering distortions to other
phases, including ferroelectric ones, may occur.

The electronics industry's demands for smaller components have made
thin ferroelectric films the subject of recent
attention.\cite{Ahn2004S,Dawber2005}  Experimentally, it is found
that the properties of ferroelectrics in thin-film form generally
differ significantly from those in the bulk.  While many factors
are expected to contribute to these differences, it has been shown
that the properties of perovskite thin films are strongly
influenced by the magnitude of the epitaxial strain resulting from
lattice-matching the film to the substrate, known as misfit strain
or epitaxial strain.

Previous theoretical studies have isolated the effects of epitaxial
strain on the structure and properties of films by imposing the
epitaxial constraint on the in-plane lattice vectors of a periodic bulk
sample.  Using a phenomenological Landau-Devonshire model, Pertsev,
Zembilgotov and Tagantsev\cite{Pertsev1998PRL} introduced the
concept of mapping the equilibrium structure of a ferroelectric
perovskite material versus temperature and misfit strain, thus
producing a phase diagram of the observable epitaxial phases.
Given the importance of strain in determining the
properties of these films, these diagrams have proven to be of
enormous interest to experimentalists seeking to interpret the
results of experiments on epitaxial thin films and
heterostructures.  This phenomenological approach should give
excellent results in the temperature/strain regime in which the
model parameters were fitted (usually near the bulk ferroelectric
transition) but will generally be less accurate when extrapolated
to other regimes.  In particular, Figure 1 of
Ref.~\onlinecite{Dieguez2004PRB} shows that two different sets of
parameters can give two quite different phase diagrams.
Furthermore, it is only possible to study materials for which all
the needed experimental information is available.

In previous work,\cite{Dieguez2004PRB,Antons2005PRB,Bungaro2004PRB} we
have examined the effects of epitaxial strain with an analogous,
but fully first-principles, approach. Specifically, we presented
density-functional theory (DFT)\cite{Hohenberg1964PR}
calculations for the structure and properties
of BaTiO$_3$, PbTiO$_3$ and SrTiO$_3$ with varying in-plane strain,
fully relaxing all structural degrees of freedom consistent with
uniform distortions (that is, retaining the five-atom unit cell).
From this, we obtained zero-temperature phase diagrams that
complement the phenomenological results of Pertsev and coworkers. In this
paper, we show that these phase diagrams can be reproduced using
the first-principles energy parameterization of King-Smith and
Vanderbilt, and give results for an additional five perovskite
oxides: CaTiO$_3$, KNbO$_3$, NaNbO$_3$, PbZrO$_3$, and BaZrO$_3$.
This approach greatly decreases the computational effort involved
in computing the phase diagram and readily allows the inclusion of
nonzero external stress.\cite{Emelyanov2002PRB} Moreover, the
parameters can be fully specified in a compact table, and the
functional form of the energy is suitable for analytical
computations and conceptual interpretation, leading to a
classification of possible stress-strain phase diagrams.
These results can be used for
predictions of the structure of epitaxially strained thin films
grown on substrates with square symmetry, and for the design of
novel perovskite strained-layer superlattices with two or more
components.

The paper is organized as follows.  In
Sec.~\ref{sec_method} we describe the extension of the KSV
method to study the effects of the epitaxial strain constraint and
external stress, and the first-principles calculations used to
obtain the KSV parameters.  Section~\ref{sec_results} presents the
results for the eight perovskite oxides considered, including the
sequence of phase transitions with varying misfit strain at zero
stress, and phase diagrams in which we show the most stable phase for
given misfit strain and external stress.  In
Sec.~\ref{sec_discussion}, we review the approximations and
discuss how to use the present results for the prediction of the
structures and polarization of thin films and superlattices.
Finally, in Sec.~\ref{sec_summary} we present our conclusions.


\section{Method}
\label{sec_method}

\subsection{Formalism}

The starting point of this analysis is the parameterized
total-energy expression presented by King-Smith and Vanderbilt in
Ref.~\onlinecite{KingSmith1994PRB}.  This is a Taylor expansion
around the cubic perovskite structure in terms of the six independent
components $\eta_i$ of the strain tensor ($i$ is a Voigt index,
$i=$1-6) and the three Cartesian components $u_{\alpha}$
(${\alpha} = x,y,z$) describing the amplitude of the soft mode
defined by the pattern of eigen-displacements associated with the
smallest eigenvalue of the (zone-center) force-constant matrix.
The arrows in
Fig.~\ref{fig_struct} indicate a typical displacement pattern
associated with this mode, greatly magnified to allow its
visualization.

The contributions to the energy (per unit cell) can divided into terms
arising from pure strain and from pure soft-mode amplitude, and an
interaction term,
    \begin{equation}
    E = E^{\rm elas}(\{\eta_i\}) + E^{\rm soft}(\{u_{\alpha}\})
      + E^{\rm int}(\{\eta_i\},\{u_{\alpha}\}) ,
    \label{eq:energy}
    \end{equation}
with the zero of the energy corresponding to the cubic structure.
For crystals with cubic symmetry the strain energy is given,
correct to second order in the strains, by
    \begin{eqnarray}
    E^{\rm elas}(\{\eta_i\})
      & = & \frac{1}{2} B_{11}
            ( \eta_1^2 + \eta_2^2 + \eta_3^2 )
            \nonumber \\
      &   & + B_{12}
            ( \eta_1 \eta_2 + \eta_2 \eta_3
            + \eta_3 \eta_1 )
            \nonumber \\
      &   & + \frac{1}{2} B_{44}
            ( \eta_4^2 + \eta_5^2 + \eta_6^2 ) ,
    \label{eq:elas}
    \end{eqnarray}
where $B_{11}$, $B_{12}$, and $B_{44}$ are related to the elastic
constants of the crystal by factors of the cell volume.  The
soft-mode energy given in Ref. \onlinecite{KingSmith1994PRB}
contains terms up to fourth-order in the soft-mode amplitude,
    \begin{equation}
    E^{\rm soft}(\{u_{\alpha}\})
      = \kappa u^2 + \alpha u^4 + \gamma ( u_x^2 u_y^2 + u_y^2 u_z^2
                                          + u_z^2 u_x^2 ) ,
    \label{eq:soft}
    \end{equation}
where $u^2 = u_x^2 + u_y^2 + u_z^2$, $\kappa$ is twice the
soft-mode eigenvalue, and $\alpha$ and $\gamma$ are the two
independent symmetry-allowed fourth-order coefficients describing the
cubic anisotropy.  Finally, the interaction between the strains and
the soft-mode amplitude is given by
    \begin{eqnarray}
    E^{\rm int}(\{\eta_i\},\{u_{\alpha}\})
      & = & \frac{1}{2} B_{1xx} ( \eta_1 u_x^2 + \eta_2 u_y^2 + \eta_3 u_z^2 )
          \nonumber \\
      &   & + \frac{1}{2} B_{1yy} [ \eta_1 ( u_y^2 + u_z^2 )
                                  + \eta_2 ( u_z^2 + u_x^2 )
          \nonumber \\
      &   &  \quad \quad \quad \quad
                                  + \eta_3 ( u_x^2 + u_y^2 ) ]
          \nonumber \\
      &   & \hspace{-1cm}
          + B_{4yz} ( \eta_4 u_y u_z + \eta_5 u_z u_x + \eta_6 u_x u_y ) ,
    \label{eq:int}
    \end{eqnarray}
where $B_{1xx}$, $B_{1yy}$, and $B_{4yz}$ are the phonon-strain
interaction coefficients. All the coefficients in these three parts
of the total-energy expression can be obtained from
first-principles calculations on a series of distorted structures
as described in Ref.~\onlinecite{KingSmith1994PRB} and in the next
subsection.

In this paper we will be concerned with the effects of strain on a
film grown epitaxially on a substrate with square symmetry.  The
epitaxial strain constraint imposed by the substrate is
    \begin{gather}
    \eta_1 = \eta_2 = \bar{\eta} , \\
    \eta_6 = 0 ,
    \end{gather}
where $\be$ is the misfit strain between the minimum-energy cubic
structure of the film material and the substrate.

In the case of epitaxy, where strain elements $\eta_1$, $\eta_2$
and $\eta_6$ are constrained while the others are not, it
is useful to introduce a mixed stress-strain elastic enthalpy
$G=E-\sigma_3\eta_3-\sigma_4\eta_4-\sigma_5\eta_5$
whose natural variables are
$u_x,u_y,u_z,\eta_1,\eta_2,\eta_6,\sigma_3,\sigma_4$ and $\sigma_5$.
Specializing to our case in which
$\eta_1=\eta_2=\bar{\eta}$, $\eta_6=0$, and assuming that the shear
stresses $\sigma_4$ and $\sigma_5$ vanish, we define an effective
elastic enthalpy given by
    \begin{equation}
    \tilde{G} = E - \sigma_3 \eta_3
    \label{eq:G}
    \end{equation}
whose natural variables are $u_x,u_y,u_z,\bar{\eta}$ and $\sigma_3$.
Using Eqs.~(\ref{eq:elas}-\ref{eq:int}) and minimizing Eq.~(\ref{eq:G})
with respect to $\eta_3$, $\eta_4$ and $\eta_5$ yields
    \begin{eqnarray}
    \eta_3 &=& \frac{1}{B_{11}} [ \sigma_3 - 2 B_{12} \be \nonumber \\
           & &                  - \frac{1}{2} B_{1xx} u_z^2
                                - \frac{1}{2} B_{1yy} (u_x^2+u_y^2) ]  \\
    \eta_4 &=& - \frac{B_{4yz}}{B_{44}} u_y u_z , \\
    \eta_5 &=& - \frac{B_{4yz}}{B_{44}} u_z u_x . 
    \end{eqnarray}
Substituting these expressions back into Eq.~(\ref{eq:G}), we express
$\tilde{G}$ in terms of its natural variables as
    \begin{eqnarray}
    \tilde G & = & (A_{\be\be} \be^2 + A_{\be\sg} \be \sg + A_{\sg \sg} \sg^2)
    \nonumber \\
    &   & + ( B_{\be} \be + B_{\sg} \sg + B ) u_{xy}^2
    \nonumber \\
    &   & + ( C_{\be} \be + C_{\sg} \sg + C ) u_{z}^2
    \nonumber \\
    &   & + D u_{xy}^4 + E u_{z}^4 + F u_{xy}^2 u_{z}^2
    \nonumber \\
    &   & + H u_{xy}^4 \sin^2\theta \cos^2\theta .
    \label{eq_gtilde}
    \end{eqnarray}
Here we have simplified the notation by replacing $\sigma_3$ by
$\sigma$. Also, the two soft-mode amplitude components are
represented in polar coordinates as
    \begin{eqnarray}
    u_x & = & u_{xy} \cos\theta , \\
    u_y & = & u_{xy} \sin\theta .
    \end{eqnarray}
The coefficients in $\tilde G$ are expressed in terms of the KSV
parameters as follows:
    \begin{eqnarray}
    &&A_{\be\be}  =   B_{11} + B_{12} - 2 \frac{B_{12}^2}{B_{11}} ,
    \\
    &&A_{\be\sg}  =   2 \frac{B_{12}}{B_{11}} ,
    \\
    &&A_{\sg\sg}  =   - \frac{1}{2 B_{11}} ,
    \\
    &&B_{\be}     =   \frac{B_{1xx}+B_{1yy}}{2}-\frac{B_{12}}{B_{11}} B_{1yy} ,
    \\
    &&B_{\sg}     =   \frac{B_{1yy}}{2 B_{11}} ,
    \\
    &&B           =   \kappa ,
    \\
    &&C_{\be}     =   B_{1yy} - \frac{B_{12}}{B_{11}} B_{1xx} ,
    \\
    &&C_{\sg}     =   \frac{B_{1xx}}{2 B_{11}} ,
    \\
    &&C           =   \kappa ,
    \\
    &&D           =   \alpha - \frac{1}{8} \frac{B_{1yy}^2}{B_{11}} ,
    \\
    &&E           =   \alpha - \frac{1}{8} \frac{B_{1xx}^2}{B_{11}} ,
    \\
    &&F           =   2 \alpha + \gamma
                      - \frac{1}{4} \frac{B_{1xx}B_{1yy}}{B_{11}}
                      -\frac{B_{1xx}^2}{B_{11}} ,
    \\
    &&H           =   \gamma .
    \end{eqnarray}

For a given set of coefficients in the potential $\tilde G$ of
Eq.~(\ref{eq_gtilde}), we can predict the phase diagram as a
function of misfit strain $\be$ and the normal external stress
$\sigma$ by minimizing $\tilde G$ to find
the values of the ground-state soft-mode amplitude components.  For
a fourth order theory, like the present KSV expression, the entire
optimization process can be done analytically, since it is possible
to compute first and second derivatives of $\tilde{G}$ and to do a
stability analysis of the various possible phases, classified by
the nature of the minimum-energy soft-mode vector.  For example, a
paraelectric $p$ phase similar to the cubic phase can appear if the
potential is minimized for ${\bf u} = 0$, but
relaxation along $\hat z$ will occur, making the cell tetragonal,
as shown in Fig~\ref{fig_struct}(b).  The classification, following
Pertsev and coworkers,\cite{Pertsev1998PRL} is given in table
\ref{tab_phases}.  Expressions for the elastic enthalpy of a given phase
as a function of misfit strain  can be obtained by minimizing
Eq.~(\ref{eq_gtilde}) with the appropriate constraint on ${\bf u}$.
For example, the elastic enthalpies for the $p$, $c$, $a$ and $aa$ phases are
\begin{eqnarray}
\tilde{G}_p    & = &  A_{\be\be} \be^2 + A_{\be\sg} \be \sg
                                       + A_{\sg \sg} \sg^2, \\
\tilde{G}_c    & = &  \tilde{G}_p - \frac{(C+C_\be \be+C_{\sg} \sg)^2}
                                         {4E}, \\
\tilde{G}_a    & = &  \tilde{G}_p - \frac{(B+B_\be \be+B_{\sg} \sg)^2}
                                         {4D}, \\
\tilde{G}_{aa} & = &  \tilde{G}_p - \frac{(B+B_\be \be+B_{\sg} \sg)^2}
                                         {4D+H}.
\end{eqnarray}

\begin{table}
\caption{Summary of epitaxial perovskite phases.
         Columns list, respectively, the phase, space group, and
         symmetry of the soft-mode amplitude components.}
\begin{ruledtabular}
\begin{tabular}{ccc}
 Phase    & SG      & SMA components  \\
\hline
 {\em p}  & $P4mmm$ &  $ u_x = u_y = u_z = 0 $                \\
 {\em c}  & $P4mm$  &  $ u_x = u_y = 0, u_z \neq 0 $          \\
 {\em a}  & $Pmm2$  &  $ u_x \neq 0 , u_y = u_z = 0 $         \\
 {\em aa} & $Amm2$  &  $ u_x = u_y \neq 0 , u_z = 0 $         \\
 {\em ac} & $Pm$    &  $ u_x \neq 0 , u_y = 0 , u_z \neq 0 $  \\
 {\em r}  & $Cm$    &  $ u_x = u_y \neq 0 , u_z \neq 0 $      \\
\end{tabular}
\end{ruledtabular}
\label{tab_phases}
\end{table}

The process of generating the stress-strain diagrams is thus
extremely rapid in comparison with a full DFT analysis, once the
KSV parameters have been obtained from first principles
calculations as described in the next section. Justification of the
approximations involved has been presented in
Ref.~\onlinecite{KingSmith1994PRB}, and will be discussed further
in Sec.~\ref{sec_discussion}.


\subsection{First-principles calculations of the coefficients}

In Table V of their paper,\cite{KingSmith1994PRB} King-Smith and
Vanderbilt report the computed coefficients to be used
in their model.  We have now repeated calculations analogous to
theirs, taking advantage of the increase in computational power
that has taken place during the last ten years to push the
boundaries of the numerical approximations that control the accuracy
of the first-principles calculations.  We used the same
local-density approximation\cite{Kohn1965PR,Ceperley1980PRB} DFT
methodology and the same ultrasoft pseudopotentials that they used.
\cite{Vanderbilt1990PRB} In our case, the plane-wave
kinetic-energy cutoff has been raised to 50 Ry, and the
Monkhorst-Pack\cite{Monkhorst1979PRB} k-point mesh is
finer, containing $8 \times 8 \times 8$ points.

The lattice parameters and the soft-mode eigenvector components
calculated for the eight perovskites under study are reported in
Table \ref{tab_eigvec}.  These values are quite similar to those
reported in Tables IV and VII of Ref.~\onlinecite{KingSmith1994PRB},
where a comparison with experimental and other theoretical data can
also be found.

\begin{table}
\caption{Lattice parameters (in bohr) and soft-mode eigenvector components
         (normalized to unity) for eight perovskites, calculated using DFT.}
\begin{ruledtabular}
\begin{tabular}{cccccc}
           &  $a_0$  &  $\xi_{\rm A}$    &  $\xi_{\rm B}$
                     &  $\xi_{\rm O_{1,2}}$  
                     &  $\xi_{\rm O_3}$  \\
\hline
BaTiO$_3$  &  7.448  &  0.184  &  0.774  & $-$0.218  & $-$0.522 \\
SrTiO$_3$  &  7.285  &  0.490  &  0.596  & $-$0.408  & $-$0.270 \\
CaTiO$_3$  &  7.201  &  0.678  &  0.363  & $-$0.435  & $-$0.171 \\
KNbO$_3$   &  7.470  &  0.195  &  0.796  & $-$0.325  & $-$0.341 \\
NaNbO$_3$  &  7.395  &  0.427  &  0.638  & $-$0.428  & $-$0.209 \\
PbTiO$_3$  &  7.337  &  0.597  &  0.487  & $-$0.411  & $-$0.262 \\
PbZrO$_3$  &  7.762  &  0.769  &  0.152  & $-$0.438  & $-$0.044 \\
BaZrO$_3$  &  7.846  &  0.703  &  0.275  & $-$0.462  & $-$0.054 \\
\end{tabular}
\end{ruledtabular}
\label{tab_eigvec}
\end{table}

Following the prescription given by King-Smith and
Vanderbilt,\cite{KingSmith1994PRB} we found the updated KSV
parameters displayed in Table~\ref{tab_coefficients}.  The two
final columns of this table contain the values of $\alpha'$ and
$\gamma'$, the coupling constants whose values determine the
symmetry of the low-temperature phase for bulk perovskites
\cite{gammaprime} (see Ref.~\onlinecite{KingSmith1994PRB}).  Most
of the parameters are only slightly different from those given in
Table V of Ref.~\onlinecite{KingSmith1994PRB}, and we have found
that the predictions that both sets give for the epitaxial films
are qualitatively the same (note, however, that cubic SrTiO$_3$
changes from being marginally unstable to marginally stable).  The
differences are mainly related to the use of finer grids in our
case when performing the fast Fourier transforms required in the
calculations.  On the other hand, the improvements in plane-wave
cutoff and k-point mesh have a quite small impact on
the values of the coefficients.

\begin{table*}
\caption{ Coefficients of energy expansion,
          Eqs.~(\protect\ref{eq:energy}-\protect\ref{eq:int}),
          for eight perovskites, in atomic units.
          Coefficients $\alpha'$ and $\gamma'$ are not directly relevant for
          this work, but are included for completeness (see
          Ref.~\protect\onlinecite{KingSmith1994PRB}).}
\begin{ruledtabular}
\begin{tabular}{lccccccccccccc}
           & $B_{11}$ & $B_{12}$ & $B_{44}$ & $B_{1xx}$ & $B_{1yy}$ & $B_{4yz}$
           & $\kappa$  & $\alpha$  & $\gamma$
           & $\alpha'$ & $\gamma'$
           \\
\hline
BaTiO$_3$
      &  4.62    &  1.80    &  1.79    &  $-$2.21    &  $-$0.19    &  $-$0.10
      &  $-$0.0143 &  0.347   &  $-$0.488
      &  0.192   &  $-$0.129
           \\
SrTiO$_3$
      &  4.97    &  1.50    &  1.54    &  $-$1.28    &   0.06    &  $-$0.12
      &   0.0010 &  0.124   &  $-$0.161
      &  0.074   &  $-$0.036
           \\
CaTiO$_3$
      &  5.34    &  1.24    &  1.41    &  $-$0.65    &   0.09    &  $-$0.11
      &  $-$0.0070 &  0.031   &  $-$0.017
      &  0.019   &  0.012
           \\
KNbO$_3$
      &  6.66    &  1.03    &  1.46    &  $-$2.91    &   0.37    &  $-$0.03
      &  $-$0.0155 &  0.439   &  $-$0.655
      &  0.257   &  $-$0.178
           \\
NaNbO$_3$
      &  6.15    &  1.18    &  0.98    &  $-$1.71    &   0.51    &  $-$0.02
      &  $-$0.0130 &  0.209   &  $-$0.298
      &  0.124   &  $-$0.050
           \\
PbTiO$_3$
      &  4.58    &  1.86    &  1.40    &  $-$0.79    &  $-$0.05    &  $-$0.07
      &  $-$0.0117 &  0.052   &  $-$0.020
      &  0.031   &  0.029
           \\
PbZrO$_3$
      &  5.34    &  1.64    &  0.93    &  $-$0.20    &   0.06    &   0.01
      &  $-$0.0185 &  0.013   &  $-$0.013
      &  0.011   &  $-$0.008
           \\
BaZrO$_3$
      &  5.69    &  1.46    &  1.45    &  $-$0.44    &   0.05    &  $-$0.12
      &   0.0087 &  0.013   &   0.003
      &  0.008   &  0.012
           \\
\end{tabular}
\end{ruledtabular}
\label{tab_coefficients}
\end{table*}

For our discussion of the effects of epitaxial strain on the polarization, it is useful to have an expansion not just of
the energy, but also of the polarization, in terms of the soft-mode
amplitude.  The linearized expression
\begin{equation}
P_z = \frac{e}{\Omega} z^{*} u_z
\label{eqn_polarization}
\end{equation}
is adequate for most purposes.
Here $e$ is the absolute value of the electron charge, $\Omega$ is
the unit cell volume of the perovskite, and $z^{*}$ is the Born
effective charge of the soft mode, given in terms of the soft-mode
eigenvectors of Table \ref{tab_eigvec} by
\begin{equation}
z^{*} = Z^{*}_{\rm A} \xi_{\rm A} + Z^{*}_{\rm B} \xi_{\rm B}
         + 2 Z^{*}_{\rm O_{1,2}} \xi_{\rm O_{1,2}} + Z^{*}_{\rm O_3}
           \xi_{\rm O_3}.
\label{eqn_modezstar}
\end{equation}
We take the Born effective charges $Z^*$ for each atom to
have their values in the cubic structure as calculated by Zhong,
King-Smith, and Vanderbilt.\cite{Zhong1994PRL}
Using these together with the eigenvectors reported in Table \ref{tab_eigvec}
we obtain the following values for $z^*$: 
9.94 (BaTiO$_3$),
8.65 (SrTiO$_3$),
7.03 (CaTiO$_3$),
11.06 (KNbO$_3$),
9.14 (NaNbO$_3$),
9.40 (PbTiO$_3$),
6.29 (PbZrO$_3$), and
5.69 (BaZrO$_3$).
This approximate expression neglects the possible dependence of the
Born effective charges upon the strain.

\section{Results}
\label{sec_results}

\subsection{Calculations at zero external stress}

We first consider the case in which the external perpendicular stress
$\sigma$ vanishes.
Table \ref{tab_transit} shows the sequence of transitions that
occurs for each of the eight perovskites studied as the misfit
strain increases, and the values of strain at which the transitions
occur.

\begin{figure}
\centerline{\epsfig{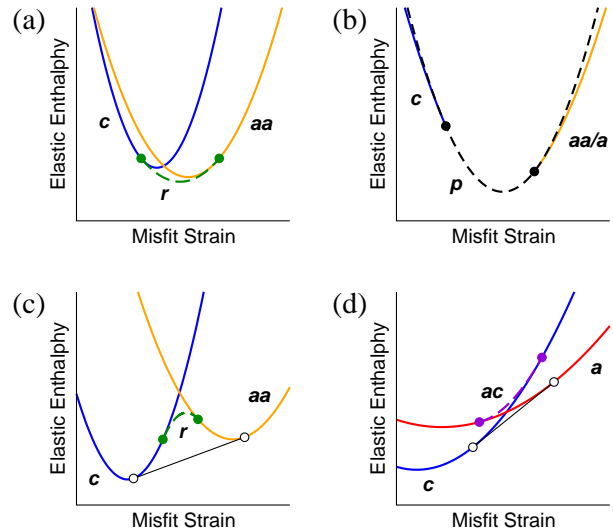}}
\caption{Sketches of different behaviors found for the elastic enthalpy
	 curves of the most stable phases.  Solid circles
	 show where two parabolas meet.  Empty circles show the
	 points where a tie line meets a parabola.  The curves that
	 meet at the circles do so with equal first derivatives.
	 }
\label{fig_parab}
\end{figure}

The observed sequences of phases can be understood with the help of
Fig.~\ref{fig_parab}, which illustrates the types of elastic enthalpy
behaviors that we observe for the materials considered.  At the strains at
which a transition occurs from one phase to another, the energy
curves join smoothly, indicating that these transitions are
of second order.  It can be shown analytically that this is indeed the
case for the KSV model, and that at the symmetry-breaking
transitions ({\em c-r, aa-r, a-ac, p-c, p-a,} or {\em p-aa}) the higher-symmetry
phase becomes unstable.  For all compounds considered, we see that
for sufficiently high compressive strains, the lowest energy phase
is always the $c$ phase, in which the atomic displacements and
therefore the polarization point in the [001] direction,
perpendicular to the substrate.  On the other hand, for
sufficiently high tensile strains, we obtain a phase in which the
polarization lies in the substrate plane, pointing along [100]
($a$ phase) for for BaZrO$_3$ or along [110] ($aa$ phase) for
the rest of the compounds.  The in-plane orientation is
determined by the sign of $H $ (i.e., of $\gamma$), which is positive for
BaZrO$_3$ and negative for the seven other compounds.  In the
intermediate strain regime, three different behaviors are found.
For BaTiO$_3$, KNbO$_3$, NaNbO$_3$, and PbZrO$_3$, an $r$ phase
appears between the $c$ and $aa$ phases, as in
Fig.~\ref{fig_parab}(a).  The fact that these perovskites crystallize in
the $r$ phase at low absolute values of strain is not surprising,
since this phase is the most similar to the rhombohedral phase that
they adopt as the bulk ground state according to the KSV theory
(see Table VI of Ref. \onlinecite{KingSmith1994PRB}).  For
SrTiO$_3$ and BaZrO$_3$ it is the paraelectric $p$ phase that
appears at intermediate strains, as in Fig.~\ref{fig_parab}(b).
For these two compounds, no $r$ or $ac$ phase appears.  Instead,
the polarization along [001] continuously goes to zero as the
strain becomes less compressive, and only reappears in the $xy$
plane once the tensile strain reaches some given value,
continuously growing from then on.  The epitaxial paraelectric $p$
phase is the analog of the bulk cubic phase (see
Fig.~\ref{fig_struct}(b)), which is the ground state predicted by
the KSV theory with the parameters of Table~\ref{tab_coefficients}
for both bulk SrTiO$_3$ and bulk BaZrO$_3$.  Finally, for CaTiO$_3$
and PbTiO$_3$ yet another behavior is obtained, as shown in
Fig.~\ref{fig_parab}(c).
At intermediate
strains, the rhombohedral phase is the lowest energy single phase.
However, partly because of its inverted-parabola elastic enthalpy curve,
the common tangent line between the $c$ and $aa$ phases yields a
lower energy in the intermediate strain regime, and thus a mixed
phase of $c$ and $aa$ domains is expected.

\begin{table}
\caption{The sequence of epitaxially-induced phase transitions
         and the values of strain $\be_{\perp}$ and
         $\be_{\parallel}$ at the boundary of the $c$ phase and of          the $aa$  (or $a$) phase, respectively.
         An asterisk denotes the strain regime where formation of mixed domains of $c$ and $aa$ phases could be favorable.}
\center
\begin{ruledtabular}
\begin{tabular}{cccc}
            &  Sequence  &  $\be_{\perp} (10^{-3})$  &  $\be_{\parallel} (10^{-3})$  \\
\hline
 BaTiO$_3$  &  {\em c-r-aa}  &   $ -5.89 $   &   $ 7.59 $ \\
 SrTiO$_3$  &  {\em c-p-aa}  &   $ -2.24 $   &   $ 1.59 $ \\
 CaTiO$_3$  &  {\em c-$\ast$-aa}  &   $ -2.30 $   &   $ 5.35 $ \\
  KNbO$_3$  &  {\em c-r-aa}  &   $ -4.80 $   &   $ 5.49 $ \\
 NaNbO$_3$  &  {\em c-r-aa}  &   $ -5.52 $   &   $ 4.13 $ \\
 PbTiO$_3$  &  {\em c-$\ast$-aa}  &   $ -3.00 $   &   $ 8.42 $ \\
 PbZrO$_3$  &  {\em c-r-aa}  &   $-52.06 $   &   $30.42 $ \\
 BaZrO$_3$  &  {\em c-p-a}   &   $-53.41 $   &   $41.86 $ \\
\end{tabular}
\end{ruledtabular}
\label{tab_transit}
\end{table}

\begin{figure}
\centerline{\epsfig{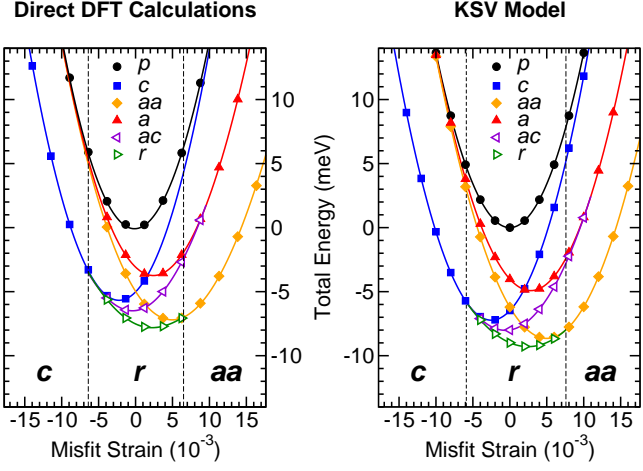}}
\caption{Comparison of BaTiO$_3$ energy curves for the six epitaxial phases
	 studied, as obtained from direct DFT
	 calculations\protect\cite{Dieguez2004PRB} (left), and from
	 the KSV theory (right). Energies are relative to the
	 paraelectric structure at zero misfit strain.  The lines
	 in the left panel and the symbols in the right panel are
	 provided as guides to the eye.}
\label{fig_btenergy}
\end{figure}

\begin{figure}
\centerline{\epsfig{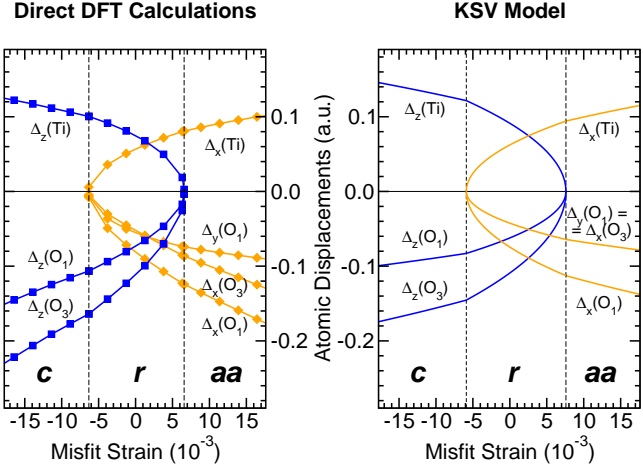}}
\caption{Comparison of BaTiO$_3$ atomic displacements for the most
	 stable phase at each given value of strain, as obtained
	 from direct DFT calculations\protect\cite{Dieguez2004PRB}
	 (left), and from our KSV model (right). 
	 $\Delta_z({\rm
	 Ti})$ indicates the displacement of the Ti atom along the
	 $z$ direction, etc.  Symmetry implies that
         $\Delta_y({\rm Ti})=\Delta_x({\rm Ti})$,
         $\Delta_x({\rm O}_2)=\Delta_y({\rm O}_1)$,
         $\Delta_y({\rm O}_2)=\Delta_x({\rm O}_1)$,
         $\Delta_z({\rm O}_2)=\Delta_z({\rm O}_1)$, and
         $\Delta_y({\rm O}_3)=\Delta_x({\rm O}_3)$.
         The lines in the left panel are guides to the eye.}
\label{fig_btdisp}
\end{figure}

The strain-induced phase transitions found using this approach
compare well with the full DFT results previously reported for
BaTiO$_3$,\cite{Dieguez2004PRB} PbTiO$_3$,\cite{Bungaro2004PRB} and
SrTiO$_3$.\cite{Antons2005PRB}
We look now in detail at this
comparison for BaTiO$_3$.  Figure \ref{fig_btenergy} shows the
energy curves of the various phases as predicted by the KSV theory
(right panel), to be compared with the full DFT results (left
panel).  The agreement between the two sets of results is very
good, with the small differences present arising from two sources.
First, the first-principles calculations in Ref.
\onlinecite{Dieguez2004PRB} were performed using the
projector-augmented wave method,\cite{Blochl1994PRB} while the
first-principles calculations used to obtain the KSV coefficients
in the present work were performed using ultrasoft pseudopotentials.
\cite{Vanderbilt1990PRB}  Second, there are the intrinsic errors
due to the use of a Taylor expansion described in the previous
section, which are expected to grow as the strain and soft mode
magnitudes increase.

Figure \ref{fig_btdisp} shows the displacements of the atoms from
their centrosymmetric perovskite positions as strain varies.
Again, the agreement of the KSV results (right panel) with the full
DFT results (left panel) is very good.  In particular, the square
root behavior predicted by the KSV theory is exhibited by the more
exact DFT calculations.  As the in-plane strain increases, we
observe a second-order phase transition ({\em c}$\text{-}${\em
r}), and while the magnitude of the atomic displacements continues
to diminish along [001], the displacements in the {\em xy} plane
begin to grow.  With increasing tensile strain, the displacements
along [001] vanish at the {\em r}$\text{-}${\em aa} transition,
while the displacements in the {\em xy} plane continue to grow
smoothly.  In this way, we see that the polarization vector
continuously rotates in going from the $c$ phase through the $r$ phase
to the $aa$ phase.  A quantitative limitation of using a single
misfit-strain-independent local mode in the KSV model is shown here
in the form of an artificial constraint equating the
$\Delta_y({\rm O}_1)$ and $\Delta_x({\rm O}_3)$ displacements.  This constraint
is removed when full DFT calculations are performed and the atoms
are free to relax within the given space group, but the magnitude
of these displacements is not very different from that obtained in
the KSV theory.

\begin{figure}
\centerline{\epsfig{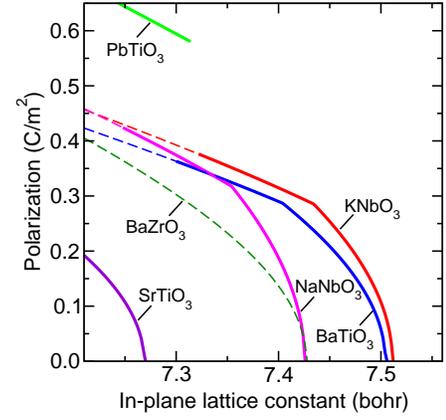}}
\caption{Value of the polarization along the $z$ axis for different perovskites.
         The continuous thick part of each curve indicates easily achievable 
         misfit strain conditions, with misfit strain
         values between $-20 \times 10^{-3}$ and $20 \times 10^{-3}$, while
         the thin dashed part of each curve corresponds to larger strains.
         The curves for CaTiO$_3$ and PbZrO$_3$ fall outside the plotted
	 region.}
\label{fig_pol}
\end{figure}

Figure~\ref{fig_pol} shows the values of $P_z$ for the various
compounds as a function of misfit strain.  The square-root
singularity at $P_z$=0 corresponds to an {\em r}$\text{-}${\em aa}
or {\em c}$\text{-}${\em p} transition.  The slope discontinuity
visible at finite $P_z$ in some curves corresponds to the {\em
c}$\text{-}${\em r} transition of Fig.~\ref{fig_parab}(a), while
the termination of the curves at finite $P_z$ for CaTiO$_3$ and
PbTiO$_3$ corresponds to the encounter with the tie line in
Fig.~\ref{fig_parab}(b).  The general trend, of course, is a strong
increase of $P_z$ with compressive strain, with considerable
enhancement possible over the zero misfit strain value.
Interestingly, according to this plot CaTiO$_3$ would
have a large $P_z$ if the zone boundary distortions that are
present in its actual ground-state crystal structure were
suppressed. This result is supported by recent full DFT
calculations.\cite{Nakhmanson2005}

\subsection{Stress-strain phase diagrams}

\begin{figure}
\centerline{\epsfig{file=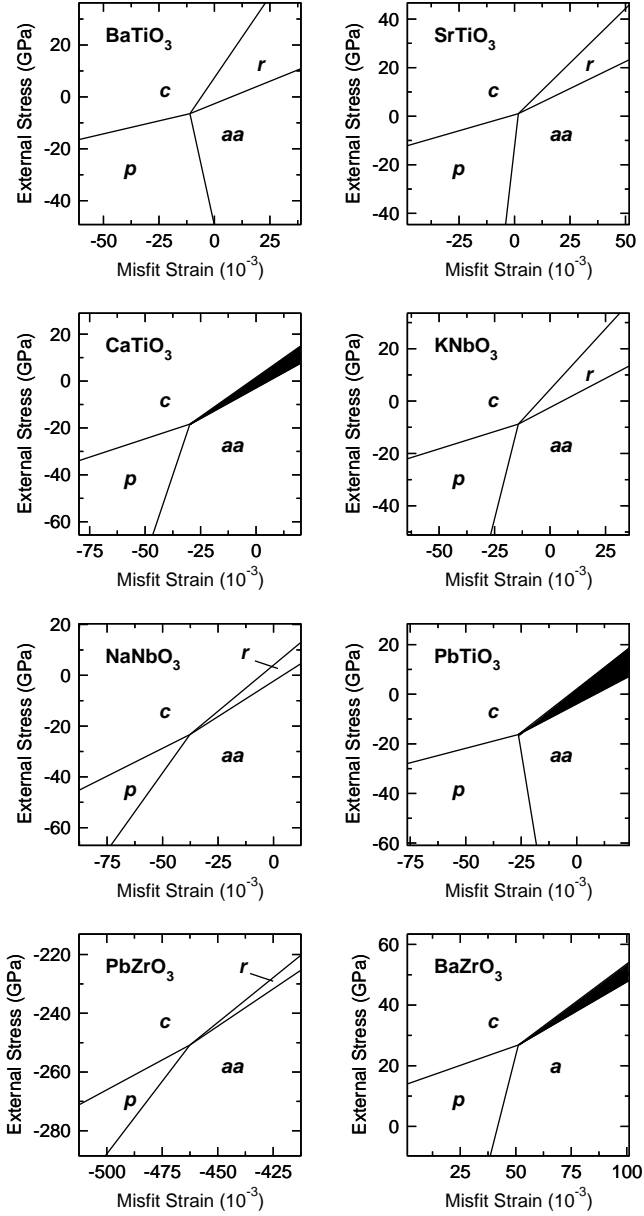,width=3.35in}}
\caption{External stress versus misfit strain phase diagrams for eight
         epitaxially strained perovskites.
         Straight lines represent second-order phase transitions.
         Shadowed areas indicate the presence of $c$ and $aa$ (or $a$)
         domains.}
\label{fig_diags}
\end{figure}

We now consider application of a nonzero normal stress $\sigma$.
The stress-strain phase diagrams obtained for each of the eight
perovskites are shown in Fig.~\ref{fig_diags}.

All eight diagrams show a universal topology with straight-line
phase boundaries meeting at a single crossing point.
The perfect linearity of the boundaries is an artifact of the
truncation of our energy expansion to fourth order, but the
presence of a single crossing point is robust against the
introduction of small higher-order terms.  The crossing point
strain $\be_{\times}$ and stress $\sigma_\times$ can be connected
with the critical strain and stress in the isotropic cubic
perovskite at which a structural instability first occurs as the
pressure is reduced or made more negative.
There are two main variants of the diagrams in Fig.~\ref{fig_diags},
the first with four phases ({\em c, r, p, \rm and \em aa/a}), and the
second with three phases ({\em c, p, aa/a \rm and a mixed phase region}). The
varied behavior of the zero-stress diagrams of Fig.~\ref{fig_parab}
can now be interpreted, in the context of Fig.~\ref{fig_diags},
as reflecting whether the zero-stress axis lies above, or below,
$\sigma_\times$.

More specifically, the coordinates of the crossing point can be
expressed as functions of the total-energy coefficients as
follows:
\begin{eqnarray}
\be_{\times} & = & \frac { B_{\sg} C - B C_{\sg} }
                         { B_{\be} C_{\sg} - B_{\sg} C_{\be} }
               = - \frac { 2 \kappa }
                         { B_{1xx} + 2 B_{1yy} } , \\
\sg_{\times} & = & \frac { B C_{\be} - B_{\be} C }
                         { B_{\be} C_{\sg} - B_{\sg} C_{\be} }
               = \be_{\times} (B_{11} + 2 B_{12}).
\end{eqnarray}
The strain and stress values at the crossing point are rather
modest, with the single exception of PbZrO$_3$. PbZrO$_3$ has the
lowest soft-mode eigenvalue, and therefore the most negative value
of $\kappa$ (Table \ref{tab_coefficients}).  Its $x$-polarized
soft-mode eigenvalue also changes more slowly on application of an
$\eta_1$ strain, resulting in a less negative value of $B_{1xx}$.
These values of $\kappa$ and $B_{1xx}$ result in $\be_{\times}$ and
$\sg_{\times}$ being about an order of magnitude more negative for
PbZrO$_3$ than for the other seven compounds.

For external stresses below $\sg_{\times}$, the behavior is similar
in all eight compounds, showing a $c$-$p$-$aa$ ($c$-$p$-$a$ for
BaZrO$_3$) sequence of second-order phase transitions, with the
elastic enthalpies of the phases behaving as in Fig.~\ref{fig_parab}(b).
In this fourth-order KSV theory, the phase boundaries are straight
lines given by
\begin{eqnarray}
\sg_{c \text{-} p}  & = & - \frac{C}{C_{\sg}} - \frac{C_{\be}}{C_{\sg}} \be
, \\
\sg_{p \text{-} aa} & = & - \frac{B}{B_{\sg}} - \frac{B_{\be}}{B_{\sg}} \be
.
\end{eqnarray}
Rewriting these expressions as functions of the fundamental
coefficients of Table \ref{tab_coefficients}, we see that the value
of $B_{1yy}$ plays a central role. 
For materials like the ones we are studying, for which $B_{11} > 0$, 
$B_{12} > 0$, and $B_{1xx} < 0$, the slope of the $c \text{-}
p$ transition line is found to be positive if 
\begin{equation}
\frac{B_{12}}{B_{11}} > \frac{B_{1yy}}{B_{1xx}} .
\end{equation}
For all eight perovskites this slope is indeed positive,
since the left hand side of the inequality is positive, 
and $B_{1yy}$ is positive or only slightly negative but
much smaller in magnitude than $B_{1xx}$ (the latter being the case
for BaTiO$_3$ and PbTiO$_3$).  The slope of the $p \text{-} aa$
transition is positive if
\begin{equation}
\frac{B_{12}}{B_{11}} > \frac{1}{2} \left( 1 + \frac{B_{1xx}}{B_{1yy}} \right),
\end{equation}
where the inequality is not satisfied for BaTiO$_3$ and PbTiO$_3$,
which have negative values for $B_{1yy}$.
Therefore, for these two perovskites a transition from
the $aa$ to the paraelectric $p$ phase is induced by applying a
sufficiently high external tensile stress at fixed misfit strain,
while for the others the transition would be from the $p$ to the
$aa$ (or $a$, in the case of BaZrO$_3$) phase.

For external stresses above $\sg_{\times}$, two kinds of behaviors
are found.  Five of the perovskites (BaTiO$_3$, SrTiO$_3$,
KNbO$_3$, NaNbO$_3$, and PbZrO$_3$) show a $c$-$r$-$aa$ sequence of
second-order phase transitions under these conditions, with the
energies of the phases behaving as in Fig.~\ref{fig_parab}(a).  In
this case, the phase boundaries are straight lines of the form
\begin{eqnarray}
\sg_{c \text{-} r}  & = & - \frac{2 B E - C F}
                        {2 B_{\sg} E - C_{\sg} F} \nonumber \\
           &   & \quad
                 - \frac{2 B_{\be} E - C_{\be} F}
                        {2 B_{\sg} E - C_{\sg} F} \be, \\
\sg_{r \text{-} aa} & = & - \frac{2 C (D+H/4) - B F}
                        {2 C_{\sg} (D+H/4) - B_{\sg} F} \nonumber \\
           &   & \quad
                 - \frac{2 C_{\be} (D+H/4) - B_{\be} F}
                        {2 C_{\sg} (D+H/4) - B_{\sg} F} \be.
\end{eqnarray}
For all five compounds, the slopes of both boundaries are
positive.

The other behavior observed for external stress above
$\sg_{\times}$ is one in which $c$ and $aa$ (or $a$, in the case of
BaZrO$_3$) domains are expected.  The energy curves behave either
as shown in Fig.~\ref{fig_parab}(c) (for CaTiO$_3$ and PbTiO$_3$)
or as shown in Fig.~\ref{fig_parab}(d) (for BaZrO$_3$).
However, in the intermediate region, instead of a
uniform phase, the system is expected to break into domains, as
explained in the previous section for PbTiO$_3$ and CaTiO$_3$ at
zero external stress.

\section{Discussion}
\label{sec_discussion}

In this section, we first consider in detail the several
approximations that are responsible for the ease and simplicity
with which we can generate stress-strain phase diagrams for
perovskites.  We then discuss and give examples of the applicability
of the theory to realistic experimental studies of perovskite
films and superlattices.

Within the KSV theory the
thermodynamical potential is expanded as
a Taylor series in strain and soft-mode amplitude,
where the reference used is the perfect cubic
perovskite of Fig.~\ref{fig_struct}(a).  The truncation at low
order in the variables of the expansion means that the expansion
decreases in accuracy for large distortions. As relevant misfit
strains are generally rather small (less than 2\%), this appears
not to have significant implications.  In addition, in the present
form of the theory, the modes selected for the expansion allow only
phases that involve five-atom unit cells to be considered.  In
particular, we do not take into account the possibility of
cell-doubling oxygen octahedra rotations, which have been shown to
be important in SrTiO$_3$,\cite{Zhong1996PRB,Vanderbilt1998F} 
CaTiO$_3$,\cite{Vanderbilt1998F} and PbZrO$_3$.\cite{Singh1997F}
In principle, such rotations could condense in the other compounds under
high enough misfit strains, but for BaTiO$_3$ it has been
shown\cite{Dieguez2004PRB} that this does not occur until one reaches
experimentally irrelevant strains, and we expect that this will also be
the case for most of these other compounds.
Second, our
calculations are done at zero temperature.  Extending them to
finite temperatures could in principle be done using an effective
Hamiltonian method as we did in Ref.~\onlinecite{Dieguez2004PRB}
for BaTiO$_3$.  However, this would involve designing for each
perovskite an effective Hamiltonian along the lines described by
Zhong, Vanderbilt, and Rabe,\cite{Zhong1994PRLandZhong1995PRB} and is
left for future work.  Third, our model does not include the small
effect of the zero-point motion of the ions (see
Ref.~\onlinecite{Iniguez2002PRL} for a discussion).  Finally, our
theory relies on the LDA to compute the exchange and correlation
terms in DFT.  This introduces small systematic errors in the
calculation, the most important of which is probably the error in
the equilibrium lattice constant.  However, such errors are well
understood and well characterized in perovskites, and tend to be
similar for different materials of this class, so that there is a
tendency for cancellation of errors in relative quantities such as
misfit strains (see, for example,
Ref.~\onlinecite{KingSmith1994PRB}).

The phenomenological Landau-Devonshire approach
\cite{Pertsev1998PRL} also requires approximations, which we review
here for the purposes of comparison. 
Its starting thermodynamical potential is the bulk free energy expanded in 
polarization and stress, with linear temperature dependence in selected 
coefficients. 
Sixth order terms are needed as their importance increases at finite 
temperature.\cite{Iniguez2001}
The reference used is the paraelectric cubic perovskite phase at the bulk 
critical temperature $T_{\text c}$, and the parameters are fit to reproduce 
experimental observations of the behavior near the bulk ferroelectric
transition. 
For the epitaxial strain dependence, a Legendre transformation is then made to 
obtain the potential as a function of polarization and misfit strain. 
With parameters extrapolated to zero temperature, this can be compared to the 
potential in the present work using the linear relation between the 
polarization and the soft mode amplitude (\ref{eqn_polarization}).
Due to the way in which the parameters are fit, the Landau-Devonshire
 potential will give its most accurate results for small misfit strains and 
temperatures near the bulk $T_{\text c}$, while the first-principles potential
will be more reliable for the zero-temperature misfit-strain phase diagram.

We now turn to the applicability of our theory to the prediction
and understanding of properties of experimentally relevant
systems.  We first discuss the case of thin films, and then consider
the case of strained-layer superlattices.

The theory presented here can be used directly to predict the
structure and polarization of a
single-domain perovskite-oxide thin film grown on a substrate with
square-lattice symmetry.  The effects of epitaxial strain will
be most evident in films coherent with the substrate. 
In equilibrium, coherent epitaxial
growth is possible up to a certain critical thickness,
which depends upon the misfit between film and substrate materials
as well as upon temperature and other growth conditions.  
Using low-temperature synthesis, coherence can be maintained far beyond the equilibrium critical thickness, as has been shown for BaTiO$_3$ grown on GdScO$_3$ ($-1.0$\% misfit strain).\cite{Choi2004S} Under such coherent conditions,
the full misfit should be used as the input for our theory.
For example, in the case of BaTiO$_3$ on GdScO$_3$,
Fig.~\ref{fig_diags} shows a predicted enhancement of $P_z$
from 0.21 to 0.31 C/m$^2$
as a result of a 1.0\% reduction of the
in-plane lattice constant from the theoretical ground-state value
of 7.448 bohr to 7.374 bohr.  For thicker,
partially-relaxed films, the strain should be taken to correspond
to the in-plane lattice constant measured for the film \cite{Speck1994JAP}; this assumes that all
the misfit dislocations responsible for the relaxation
are located at the interface.
These predictions are based on the assumption that the epitaxial strain
strongly dominates other factors in determining the state of the
film.

Similarly, for strained-layer superlattices, the
structure and polarization of a perovskite-oxide layer in a
superlattice will in general be significantly different from the
corresponding bulk.  
The states of the component layers will
largely be determined by the in-plane strain imposed by lattice
matching to the other components and/or to the substrate, and
can be obtained by referring to the theoretical phase diagrams at
the relevant value of misfit strain. In addition, the normal component of the
polarization in each layer, at the common in-plane strain, is
an important consideration in determining the structure and
polarization of the overall superlattice. If the normal
polarization is discontinuous, electrostatic energy considerations
will tend to polarize the low-polarization layer and depolarize the
high-polarization layer to make the normal polarization uniform
through the sublattice,\cite{Neaton2003} with accompanying changes
in the normal strain of the layer.
As an example, this analysis has proved useful in
the interpretation of first-principles results and experiments on
the structure of partially-relaxed SrTiO$_3$/BaTiO$_3$ superlattices.
\cite{Johnston2005,Rios2003JPCM}

First-principles information about the effects of epitaxial strain on the
structure and polarization of component layers also allows us to determine the 
relative stability of a given superlattice. 
It is well known that lattice matching of the components minimizes the elastic 
energy and thus increases the stability. In the case where one or more 
components have a nonzero normal polarization, the electrostatic energy of the 
superlattice can be minimized by  ``polarization matching," where the 
components are selected to have the same normal polarization at the common 
in-plane strain. For example, according to our calculations,
BaTiO$_3$ and NaNbO$_3$ have the same normal polarization of 0.34 
C/m$^2$ at a common in-plane theoretical lattice constant of 7.336 bohr, 
corresponding to small compressive misfit strains of less that 0.2\%, and thus 
this is a favorable combination with respect both to elastic and electrostatic 
energy. In general, however, it is not possible to minimize both elastic and 
electrostatic energy in this way, and a trade-off between the two is necessary to 
form the superlattice.

It should be kept in mind, however, that for both films and superlattices, the
assumptions that the system is in a single domain and that the epitaxial 
strain strongly dominates other factors will not be valid in all cases.
Phase diagrams including multiple-domain states have, for example, been 
discussed in Refs. \onlinecite{Speck1994JAP,Pertsev2000PRL,Li2003APL}. 
Other influences that may be important include surface relaxation and
reconstruction, atomic and electronic rearrangements at the
interface, imperfectly compensated macroscopic electric fields,
deviations from stoichiometry, and the presence of defects.


\section{Summary}
\label{sec_summary}

We have applied the first-principles total-energy parameterization
of King-Smith and Vanderbilt\cite{KingSmith1994PRB} to study
the effects of epitaxial strain and external stress on the
structure and properties of perovskites.  We report phase diagrams
and polarizations for the same set of eight compounds as in Ref.
\onlinecite{KingSmith1994PRB}: BaTiO$_3$, SrTiO$_3$, CaTiO$_3$,
KNbO$_3$, NaNbO$_3$, PbTiO$_3$, PbZrO$_3$, and BaZrO$_3$. An
updated set of parameters, computed with a comparable
first-principles method at higher precision, are provided in Table
\ref{tab_coefficients}. The simple form of the parameterization is
seen to be useful in reducing the computational effort for
generating these phase diagrams relative to full first-principles
calculations, and many features of the phase diagrams can be
extracted and interpreted analytically.

We have discussed the use of these results in predicting the
structure and polarization of epitaxial perovskite films and
strained layer superlattices.  Additional properties of interest,
such as dielectric or piezoelectric constants can also be computed
within this framework.


\begin{acknowledgments}
We thank Javier Junquera for useful discussions.  This work was
supported by ONR Grants N0014-05-1-0054,
N00014-00-1-0261, and
N00014-01-1-0365, and DOE Grant DE-FG02-01ER45937.
\end{acknowledgments}



\end{document}